\documentclass[twocolumn,amsmath,amssymb,aps,prb,superscriptaddress]{revtex4-2}
\usepackage{bm}
\usepackage{xcolor}
\usepackage[added=red]{changes}
\definechangesauthor[name={Added}, color=red]{added}
\usepackage{ulem} 
\usepackage{times}
\usepackage{amsmath}
\usepackage{mathrsfs}
\usepackage{graphicx}
\usepackage{epstopdf}
\usepackage{amssymb}
\usepackage{natbib}
\usepackage[colorlinks=true, letterpaper=ture, pdfstartview=FitV, linkcolor=blue, citecolor=blue, urlcolor=blue]{hyperref}
\usepackage{dcolumn}
\usepackage{hyperref}

\renewcommand{\theequation}{\arabic{equation}}

\begin{document}
\title{The influence of magnon renormalization and interband coupling on the spin Seebeck effect in YIG}
\author{Yuling Yin}
\affiliation{National Laboratory of Solid State Microstructures and School of Physics, Nanjing University, Nanjing 210093, China}
\affiliation{Collaborative Innovation Center of Advanced Microstructures, Nanjing University, Nanjing 210093, China}
\affiliation{International Quantum Academy, Shenzhen 518048, China}

\author{Yang Liu}
\affiliation{National Laboratory of Solid State Microstructures and School of Physics, Nanjing University, Nanjing 210093, China}
\affiliation{Collaborative Innovation Center of Advanced Microstructures, Nanjing University, Nanjing 210093, China}
\affiliation{International Quantum Academy, Shenzhen 518048, China}

\author{Yiqun Liu}
\thanks{E-mail: liuyiqun@mail.bnu.edu.cn}
\affiliation{School of Physics and Electronic Science, Changsha University of Science and Technology, Changsha 410114, China}\affiliation{National Laboratory of Solid State Microstructures and School of Physics, Nanjing University, Nanjing 210093, China}
\affiliation{Collaborative Innovation Center of Advanced Microstructures, Nanjing University, Nanjing 210093, China}

\author{Xiangang Wan}
\thanks{E-mail: xgwan@nju.edu.cn}
\affiliation{National Laboratory of Solid State Microstructures and School of Physics, Nanjing University, Nanjing 210093, China}
\affiliation{Collaborative Innovation Center of Advanced Microstructures, Nanjing University, Nanjing 210093, China}
\affiliation{Hefei National Laboratory, Hefei 230088, China}
\affiliation{Jiangsu Physical Science Research Center, Nanjing 210093, China}

\date{\today}

\begin{abstract}

With exceptionally low magnetic damping, YIG has been extensively applied in the realm of magnetism, encompassing the researches into the spin Seebeck effect.
YIG has 20 magnon bands, including 8 higher-energy bands denoted as $\alpha_{1\sim8}$, and 12 lower-energy bands denoted as $\beta_{1\sim12}$.
Here, we study the impact of the complex intraband and interband magnon couplings on the transport coefficients of YIG.
Four-magnon processes in YIG are considered, and a self-consistent mean-field approximation is made for these interaction terms.
We find that the $\beta$ bands exhibit minimal variation with increasing temperature, whereas the $\alpha$ bands undergo a noticeable decline as the temperature rises.
These counterintuitive results agree well with the observation of earlier inelastic neutron scattering experiments and the results of the theoretical calculations in recent years.
We also find it sufficient to include only the contribution of magnons on the acoustic band $\beta_1$ when studying the spin conductivity ($\sigma_m$).
However, when calculating the spin Seebeck coefficient ($S_m$) and the magnon thermal conductivity ($\kappa_m$), the results calculated using only $\beta_{1}$ show noticeable deviations over a large temperature range compared to the full band calculations. These deviations are well mitigated when the $\beta_{2}$ and $\beta_{3}$ bands are considered.
Finally, we find our results satisfy the Wiedemann-Franz law for magnon transport.

\end{abstract}
\maketitle
\section{Introduction}
Spin-caloritronics is a field that explores the interaction of electron spins with heat currents, driven by new physical effects and aimed at enhancing thermoelectric device performance \cite{bauer2012spin, bauer2010spin, boona2014spin,chumak2015magnon, YU2017pla, Kikkawa2023}.
This field is anticipated to address efficient waste heat recovery and temperature control challenges \cite{Kikkawa2023}.
The comprehension of spin-caloritronics phenomena pivots on magnon transport in magnetic insulators, notably the spin Seebeck effect (SSE) and its reverse phenomenon, the Peltier effect \cite{uchida2008,uchida2010spin,kajiwara2010,Adachi2013,DanielLoss2015,Rezende2016SSE}.
SSE stems from spin currents generated by thermal gradients, encompassing key physical quantities such as magnon conductivity ($\sigma_m$), spin Seebeck coefficient ($S_m$), and magnon thermal conductivity ($\kappa_m$) \cite{Adachi2013, DanielLoss2015, Rezende2016SSE}. It has been recognized as a ubiquitous physical phenomenon in magnetic materials since its experimental confirmation in 2008 \cite{uchida2008, Kikkawa2023}.

In spin caloritronics, closely intertwined with the SSE detections, yttrium iron garnet (YIG) emerges as a prominent material choice, primarily attributed to its remarkably low magnetic damping \cite{kajiwara2010,Serga2010, Cherepanov1993saga, Barman2021, Azzawi2017, Arsad2023}. 
This low magnetic damping attribute enables spin-waves to propagate to an observable distance \cite{cornelissen2015long,wesenberg2017long,Gomez-Perez2020}.
So far, significant and concerted efforts have been made to explore the SSE in YIG, both experimentally and theoretically \cite{Arsad2023,Rezende2014, Kikkawa2015,Jin2015, Kikkawa2016,Wang2018}.

YIG has a high Curie temperature of $T_c\approx560 K$, allowing experiments to be conducted at room temperature \cite{Cherepanov1993saga}. 
The lattice of YIG belongs to the space group of $Ia\bar 3d$, with a primitive unit cell containing 80 atoms, of which 20 are iron atoms. 
Despite of complex structure, the crystal of YIG can grow extremely well, achieving acoustic damping even lower than that of quartz \cite{Cherepanov1993saga}.
YIG has 20 magnon modes, including one acoustic branch and 19 optical branches \cite{Rezende2020fundamentals}.
When the wave-vector ${\bm k}<5\times 10^5cm^{-1}$, the acoustic magnon can be probed by microwave techniques and inelastic light scattering \cite{Rezende2020fundamentals}.
However, it is necessary to use inelastic neutron scattering (INS) experiments when detecting higher-energy magnons \cite{Cherepanov1993saga, Rezende2020fundamentals,Princep2017}.
At room temperature, the wavelength of thermal magnons in YIG is a few nanometers \cite{Cornelissen2016,Boona2014}. These magnons are localized near the $\Gamma$ point of the acoustic branch, with a frequency on the order of GHz.
The magnons excitation in YIG will undergo a rapid thermalization within hundreds of nanoseconds \cite{Ruckriegel2014,Streib2019, Maehrlein2018}.
Therefore, in the realm of magnon spin transport research \cite{Serga2010,Arsad2023}, the acoustic branch of magnons has garnered significant attention, whereas the impact of the optical branches is often overlooked.
From this perspective, when the optical modes are considered, the interband coupling of magnons in YIG remains an interesting issue.
In recent years, in pursuit of higher computational frequencies, there has been a growing demand to manipulate magnons with frequencies exceeding THz \cite{chumak2015magnon,Barman2021,nikitov2020dielectric}. 
For YIG, the frequency of $\sim6$THz is sufficient to excite optical magnons \cite{Gorbatov2021,Shen2018,Xie2017,Liu2017}. 
Hence, the excitation of optical magnons and the interband coupling in YIG are destined to attain escalating significance.
Furthermore, the interband coupling could potentially account for the minimal energy variation observed in the acoustic branches of YIG as the temperature increasing \cite{Barker2016}.

In this paper, a self-consistent mean-field approximation methods are employed to determine the temperature-dependent magnon energy spectra in YIG. 
Then, leveraging the temperature-dependent energy spectrum, we investigate the temperature dependence of the transport coefficients ($\sigma_m$,  $\kappa_m$, $S_m$). Both the interband and intraband magnon couplings are included to investigate these transport coefficients.

\section{Model Hamiltonian and Formalism}\label{Sec-2}
\subsection{The quadratic magnon Hamiltonian of YIG}
Within a primitive unit cell of YIG, 20 Fe atoms are arranged in a colinear order, with 8 on the $a$ sites and 12 on the $d$ sites\cite{Cherepanov1993saga, ABHarris1963, Shen2018}. 
The spins of the $a$ and $d$ sites are oppositely aligned, forming a ferrimagnetic order.
In the construction of magnetic order, the exchange interactions play a dominant role and undergo a rapid decline as the distance increases.
In addition, compared to the exchange interactions in YIG, the magnetic dipole interactions are very small.
In YIG, the magnetic exchange interaction is on the order of 1000K, while the magnetic dipole interaction is only about 1K \cite{Cherepanov1993saga}.
Thus the magnetic dipole interaction is ignored in this paper.
In alignment with these earlier studies \cite{Cherepanov1993saga,Shen2018}, we start with a Hamiltonian encompassing three strongest Heisenberg exchange interactions alongside exceedingly minor external Zeeman energies as follows:
\begin{equation}\label{HHH}
\begin{aligned}
H=&-\sum_{n=1}^N\Big[J_{aa}\sum_{i=1}^8 \mathbf{S_a}(\mathbf{R}_{n,i})\!\cdot\!\sum_{{|{\mathbf r}_{ij}|}=\rm{r_{aa}}} \mathbf{S_a}(\mathbf{R}_{n,i}\!+\!\mathbf{r}_{ij})\\
&+J_{dd}\sum_{i=9}^{20} \mathbf{S_d}(\mathbf{R}_{n,i})\!\cdot\!\sum_{{|{\mathbf r}_{ij}|}=\rm{r_{dd}}} \mathbf{S_d}(\mathbf{R}_{n,i}\!+\!\mathbf{r}_{ij})\\
&+2J_{ad}\sum_{i=1}^8 \mathbf{S_a}(\mathbf{R}_{n,i})\!\cdot\!\sum_{{|{\mathbf r}_{ij}|}=\rm{r_{ad}}} \mathbf{S_d}(\mathbf{R}_{n,i}\!+\!\mathbf{r}_{ij})\\
&+\sum_{i=1}^8g\mu_{B}\mathbf{B}\!\cdot\! \mathbf{S_a(R}_{n,i})+\sum_{i=9}^{20}g\mu_{B}\mathbf{B}\!\cdot\! \mathbf{S_d(R}_{n,i})\Big],
\end{aligned}
\end{equation}
where $N$ represents the total number of primitive unit cells, $n$ denotes the primitive unit cell, and $i$ and $j$ are indices referring to the sublattices.
Additionally, $\rm{r_{aa}}$ denotes the shortest distance between $a$ sites, $\rm{r_{dd}}$ denotes the shortest distance between $d$ sites, and $\rm{r_{ad}}$ represents the minimal distance between $a$ sites and $d$ sites.
And $J_{aa}$, $J_{dd}$, and $J_{ad}$ denote the corresponding exchange interactions. The small external magnetic field $\mathbf{B}$ is set along the $z$-component of ${\bf S_d}$.

Utilizing Holstein-Primakoff (HP) transformation \cite{HP1940,Oguchi1960}, the spin Hamiltonian above could be described in a magnon quasiparticle form. The HP transformation for the ferrimagnet YIG is expressed as follows:
\begin{equation}\label{HP}\begin{aligned}
S_a^z&=\;S-a^\dagger a,&S_a^-&=(S_a^+)^\dagger=a^\dag\sqrt{2S-a^\dag a},\\
S_d^z&=-S+d^\dagger d,&S_d^+&=(S_d^-)^\dagger=d^\dag\sqrt{2S-d^\dag d}.
\end{aligned}\end{equation}
Thus, substituting Eq. (\ref{HP}) into Eq. (\ref{HHH}) by expanding in terms of powers of $1/S$ \cite{Oguchi1960,Igarashi1992}, the magnon Hamiltonian can be expressed as 
\begin{equation}\label{HHH-taylor}
H=H^{(0)}+H^{(2)}+H^{(4)}+\cdots,
\end{equation}
where $H^{(0)}$ represents the background energy, $H^{(2)}$ represents the kinetic energy of magnon in quadratic form, and $H^{(4)}$ represents the leading terms of magnon-magnon interactions \cite{Dyson1956, Kubo1952, HKHH1971}.
The higher-order terms represented by the ellipsis in Eq. (\ref{HHH-taylor}) are ignored.
Employing the Fourier transformations $a_{\mathbf{R}_{n,i}}=\frac1{\sqrt{N}}\sum_{\bf k}e^{i\bf k\cdot \bf{R}_{n,i}}a_{i\bf k}$ and $d_{\mathbf{R}_{n,i}}=\frac1{\sqrt{N}}\sum_{\bf k}e^{i\bf k\cdot \bf{R}_{n,i}}d_{i\bf k}$ on $H^{(2)}$, the harmonic Hamiltonian $H^{(2)}$ is expressed as
\begin{eqnarray}\label{HH2}
H^{(2)}&=&\sum_{\bf{k}} \begin{matrix}\big(\bm{a}_{\bf{k}}^\dag & \bm{d}_{-{\bf{k}}}\big)\end{matrix}\Bigg[\begin{matrix}\bm A_{\bf{k}} & \bm B_{\bf{k}}\\ \bm B_{\bf{k}} ^\dag & \bm D_{\bf{k}}\end{matrix}\Bigg]
\Bigg(\begin{matrix}\bm{a}_{\bf{k}}\\\bm{d}_{-{\bf{k}}}^\dag\end{matrix}\Bigg)\nonumber\\
&=&\sum_{\bm k}\bm{x}_{\bf{k}}^\dag \bm{\mathcal H}_{\bf{k}}\bm{x}_{\bf{k}},
\end{eqnarray}
where $\bf A_{\bf k}$, $\bf B_{\bf k}$ and $\bf D_{\bf k}$ are $8\times8$, $8\times12$ and $12\times12$ matrices, respectively, with each corresponding matrix element given by:
\begin{equation}\begin{aligned}&\bm A_{\bf{k}}^{ij}=(16J_{aa}S-12J_{ad}S+g\mu_BB)\delta_{ij}-2J_{aa}S\gamma_{\bf k}^{ij},\\
&\bm D_{\bf{k}}^{ij}=(8J_{dd}S-8J_{ad}S-g\mu_BB)\delta_{ij}-2J_{dd}S\gamma_{\bf k}^{ij},\\
&\bm B_{\bf{k}}^{ij}=-2J_{ad}S\gamma_{\bf k}^{ij}\nonumber,\end{aligned}\end{equation}
and $\begin{matrix}\big(\bm{a}_{\bf{k}}^\dag & \bm{d}_{-{\bf{k}}}\big)\end{matrix}=\begin{matrix}\big({a}_{1\bf{k}}^\dag \cdots {a}_{8\bf{k}}^\dag\; {d}_{{9-\bf{k}}}\cdots {d}_{{20-\bf{k}}}\big)\end{matrix}$. Moreover, we set $\bm{x}_{\bf{k}}^\dag=\begin{matrix}\big(\bm{a}_{\bf{k}}^\dag & \bm{d}_{-{\bf{k}}}\big)\end{matrix}$ for simplicity. 
Note that the structure factor $\gamma_{\bf k}^{ij}$ is defined as the summation over the neighboring atoms of the $i$th atom in the sublattice: 
\begin{equation}\label{gamma}\gamma_{\bf k}^{ij}=\sum_{j\in\langle i,j\rangle}{e^{i{\bf k}\cdot({\bf r}_{j}-{\bf r}_{i})}},\end{equation} 
where $\sum_{j\in\langle i,j\rangle}$  is a summation over the $j$th site which are restricted to the neighbors of the $i$th site, connected by the $J_{aa}, J_{dd}$ or $J_{ad}$ bonds. Here, ${{\bf r}_{ij}}={{\bf r}_{j}}-{{\bf r}_{i}}$ satisfies $|{{\bf r}_{ij}}|={\rm r_{aa}}$, ${\rm r_{dd}}$, or ${\rm r_{ad}}$. The exchange constants are set as $J_{ad} = -39.8 {K}$, $J_{dd} = -13.4 {K}$, and $J_{aa} = -3.8 {K}$ \cite{Cherepanov1993saga}.

In the diagonalization process \cite{WhiteRM1965, Colpa1978}, the quadratic Hamiltonian (\ref{HH2}) can be solved with
\begin{equation}\begin{aligned}\label{Diag}
\bm{x}_{\bf{k}}^\dag \bm{\mathcal H}_{\bf{k}}\bm{x}_{\bf{k}}
&= \bm{x}_{\bf{k}}^\dag {(\bm{\mathcal P}^\dag)^{-1}}{\bm{\mathcal P}^\dag} \bm{\mathcal H}_{\bf{k}}{\bm{\mathcal P}}{\bm{\mathcal P}^{-1}}\bm{x}_{\bf{k}}\\ &=\bm{x}_{\bf{k}}^\dag {(\bm{\mathcal P}^\dag)^{-1}}\bm{\mathcal E}_{\bf{k}}{\bm{\mathcal P}^{-1}}\bm{x}_{\bf{k}}=\bm{\chi}_{\bf{k}}^\dag \bm{\mathcal E}_{\bf{k}}\bm{\chi}_{\bf{k}}.
\end{aligned}\end{equation}
The transformed state $\bm{\chi}_{\bf{k}}^\dag=(\chi_{1\bf k}^\dag\cdots \chi_{8\bf k}^\dag\,\chi_{9\bf k}^\dag\cdots \chi_{20\bf k}^\dag)=\big({\alpha}_{1\bf{k}}^\dag \cdots {\alpha}_{8\bf{k}}^\dag\; {\beta}_{{1-\bf{k}}}\cdots {\beta}_{{12-\bf{k}}}\big)$ maintains the following commutation relations: 
\begin{equation}\begin{aligned}[\alpha_{i\mathbf{k}},\; \alpha_{j\mathbf{k}'}^\dagger] =& \delta_{ij}\delta_{\mathbf{k}\mathbf{k}'}, \quad [\beta_{i\mathbf{k}},\; \beta_{j\mathbf{k}'}^\dagger] = \delta_{ij}\delta_{\mathbf{k}\mathbf{k}'}, \\
[\beta_{i\mathbf{k}},\; \alpha_{j\mathbf{k}'}^\dagger] &= [\alpha_{i\mathbf{k}},\; \beta_{j\mathbf{k}'}^\dagger] = 0.\nonumber
\end{aligned}\end{equation}
Unlike Fermion systems, the diagonalization of the magnetic Hamiltonian $\bm{\mathcal H}$ to a diagonal form $ \bm{\mathcal E}={\rm Diag}\{\varepsilon_1,\varepsilon_2,\cdots,\varepsilon_{20}\}$ for bosonic systems necessitates a para-unitary transformation $\bm{\mathcal P}$ \cite{Colpa1978}. To obtain $\bm{\mathcal E}$ and $\bm{\mathcal P}$, a metric matrix ${\bf g}$ should be  introduced as follows \cite{Colpa1978, Shenka2017}:
\begin{equation}\label{metric}
\bf g=\begin{bmatrix}\bm I_{8\times8} & \bm 0 \\ \bm 0 & -\bm I_{12\times12}\end{bmatrix},
\end{equation}
where $\bm I_{8\times8}$ is an $8\times8$ identity matrix, and $\bm I_{12\times12}$ is a $12\times12$ identity matrix. The diagonalization process is divided into three sequential steps \cite{Colpa1978}. Initially,  we apply the Cholesky decomposition $\bm{\mathcal{H}= \mathcal K^\dag \mathcal K}$ to find an upper-triangular matrix $\bm{\mathcal K}$.  Subsequently, a unitary diagonalization is executed on the resultant matrix $\bm{\mathcal K}\bf{g}\bm{\mathcal K}^\dag$, yielding the energy eigenvalues matrix $\bf{g} \bm{\mathcal E}$ along with the unitary matrix $\bm{\mathcal U}$. Finally, the Bogoliubov transformation matrix $\bm{\mathcal P}$ can be computed as $\bm{\mathcal P} = \bm{{\mathcal E}^{1/2}\mathcal U{\mathcal K}^{-1}}$. As a result, the para-unitary transformation $\bm{\mathcal P}$ and the 20 magnon quadratic energies $\bm{\mathcal E}$  can be obtained. The dispersions without the contribution from $H^{(4)}$ along the $[110]$ and $[100]$ directions are depicted in Fig. \ref{YIG-e}(a). 

\subsection{Magnon-magnon interaction and mean-field approximation}
The magnon-magnon interaction term $H^{(4)}$ in Eq. (\ref{HHH-taylor}) will introduce a temperature-dependent correction to the magnon spectra of YIG \cite{Shen2018,Liu2023}. The derivation details of $H^{(4)}$ are shown in Appendix~\ref{app: A}. After Fourier transformation, $H^{(4)}$ can be given as Eq. (\ref{HH4app}) .

As suggested in Eq. (\ref{Diag}), the Hamiltonian should be expressed in terms of the eigenvectors $\chi_{m{\bf k}}$ \cite{Arakawa2022}. In terms of matrix elements, the Bogoliubov transformation can be written as: 
\begin{equation}\label{trans}
x_{i{\bf k}}=\sum_{m=1}^{20}\mathcal P_{\bf k}^{i,m}\chi_{m{\bf k}},\quad x_{i{\bf k}}^\dagger=\sum_{m=1}^{20}\left(\mathcal P_{\bf k}^{i,m}\right)^*\chi_{m{\bf k}}^\dagger.
\end{equation}
And $H^{(4)}$ in Eq. (\ref{HH4app}) should also be expressed in terms of $(\chi_{m{\bf k}},\chi_{m{\bf k}}^\dag)$, for example, $a^{\dagger}_{i\bf{k}_4}a^{\dagger}_{i\bf{k}_3}a_{i\bf{k}_2}a_{j\bf{k}_1}$ can be writted as $\sum_{m_1\sim m_4=1}^{20}$ $\mathcal{P}_{\bf{k}}^{i,m_4*}$ $\mathcal{P}_{\bf{k}}^{i,m_3*}$ $\mathcal{P}_{\bf{k}}^{i,m_2}$ $\mathcal{P}_{\bf{k}}^{j,m_1}$\ $\chi_{m_4{\bf k}_4}^\dagger\!$ $ \chi_{m_3{\bf k}_3}^\dagger\!$ $\chi_{m_2{\bf k}_2}\!$ $\chi_{m_1{\bf k}_1}\!$. The transformed four-operator terms are organized in three orderings: $\chi_{m_4{\bf k}_4}^\dagger\! \chi_{m_3{\bf k}_3}^\dagger\!\chi_{m_2{\bf k}_2}\!\chi_{m_1{\bf k}_1}$, $\chi_{m_1{\bf k}_1}\!\chi_{m_2{\bf k}_2}\! \chi_{m_3{\bf k}_3}^\dagger\!\chi_{m_4{\bf k}_4}^\dagger$, and  $\chi_{m_4{\bf k}_4}^\dagger\!\chi_{m_2{\bf k}_2}\! \chi_{m_3{\bf k}_3}^\dagger\!\chi_{m_1{\bf k}_1}\!$. We can reduce the complexity of four magnon operators to two operators using a mean-field approximation \cite{Liu2023,Bloch1962,Shen2018}, for example,
\begin{equation}\begin{aligned}\label{chichi}
\chi_{m_4{\bf k}_4}^\dagger\! \chi_{m_3{\bf k}_3}^\dagger\!\chi_{m_2{\bf k}_2}\!\chi_{m_1{\bf k}_1}\!
&\!\approx\!\langle\chi_{m_4{\bf k}_4}^\dagger\!\chi_{m_1{\bf k}_1}\rangle \chi_{m_3{\bf k}_3}^\dagger\! \chi_{m_2{\bf k}_2}\\
&\!+\!\langle\chi_{m_3{\bf k}_3}^\dagger\! \chi_{m_2{\bf k}_2}\rangle \chi_{m_4{\bf k}_4}^\dagger\! \chi_{m_1{\bf k}_1}\\
&\!+\!\langle\chi_{m_4{\bf k}_4}^\dagger\! \chi_{m_2{\bf k}_2}\rangle \chi_{m_3{\bf k}_3}^\dagger\! \chi_{m_1{\bf k}_1}\\
&\!+\!\langle\chi_{m_3{\bf k}_3}^\dagger\! \chi_{m_1{\bf k}_1}\rangle \chi_{m_4{\bf k}_4}^\dagger\! \chi_{m_2{\bf k}_2}.
\end{aligned}\end{equation}
And the Bose distribution functions $n_{m_\mu{\bf k_\mu}}$ of the $m_\mu$th bands are defined as:
\begin{equation}\begin{aligned}\label{bose}
\langle\chi_{m_\mu{\bf k_\mu}}^\dagger\chi_{m_\nu{\bf k_\nu}}\rangle= \delta_{m_\mu m_\nu}\delta_{{\bf k_\mu}{\bf k_\nu}}(n_{m_\mu{\bf k_\mu}}+\xi^{\beta}_{m_\mu}),\\
\langle\chi_{m_\mu{\bf k_\mu}}\chi_{m_\nu{\bf k_\nu}}^\dagger\rangle= \delta_{m_\mu m_\nu}\delta_{{\bf k_\mu}{\bf k_\nu}}(n_{m_\mu{\bf k_\mu}}+\xi^{\alpha}_{m_\mu}),
\end{aligned}\end{equation}
where $\xi_{m}^{\beta}=1$ when $\chi_{m\bf{k}}$ is a $\beta$ state and $\xi_{m}^{\alpha}$ =1 when $\chi_{m\bf{k}}$ is an $\alpha$ state. Otherwise, they equal zero. In our case,
\begin{equation}\begin{aligned}\label{addii}
&\langle{a}_{i\bf{k}}^\dag {a}_{i\bf{k}}\rangle =\sum_{m=1}^{20}|{\mathcal P}_{{\bf k}}^{i,m}|^2 n_{m\bf{k}}+\sum_{m=9}^{20}|{\mathcal P}_{{\bf k}}^{i,m}|^2,\\
&\langle{{d}_{j-\bf{k}}^\dag {d}_{j-\bf{k}}}\rangle =\sum_{m=1}^{20}|{\mathcal P}_{{\bf k}}^{j,m}|^2 n_{m\bf{k}}+\sum_{m=1}^{8}|{\mathcal P}_{{\bf k}}^{j,m}|^2,\\
\end{aligned}\end{equation}
and when $ i\neq j$, 
\begin{equation}\label{addij}\begin{aligned}
\langle{x}_{i\bf{k}}^\dag{x}_{j\bf{k}}\rangle &=\sum_{m=1}^{20}{\mathcal P}_{{\bf k}}^{i,m*}{\mathcal P}_{{\bf k}}^{j,m} n_{m\bf{k}}+\sum_{m=9}^{20}{\mathcal P}_{{\bf k}}^{i,m*}{\mathcal P}_{{\bf k}}^{j,m}\\
&=\sum_{m=1}^{20}{\mathcal P}_{{\bf k}}^{i,m*}{\mathcal P}_{{\bf k}}^{j,m} n_{m\bf{k}}+\sum_{m=1}^{8}{\mathcal P}_{{\bf k}}^{i,m*}{\mathcal P}_{{\bf k}}^{j,m}.
\end{aligned}\end{equation}
Substituting Eqs. (\ref{trans} -- \ref{addij}) into Eq. (\ref{HH4app}),the energy correction can be succinctly expressed as:
\begin{equation}\label{MF}
H^{(4)}\approx\sum_{\bf{k}}\sum_{m=1}^{20}\Delta\tilde\varepsilon_{m\bf{k}}\chi_{m\bf{k}}^\dagger \chi_{m\bf{k}}
\end{equation}
with
\begin{eqnarray}\label{delta}
\Delta\tilde\varepsilon_{m\bf{k}}=&&\sum_{i=1}^{20}\sum_{j\in\langle i,j\rangle}\!\!2J_{ij}\\
&&\times{\rm Re}\Big[\eta_{ij}\Big(e^{i{{\bf k}}\cdot{\bf r}_{ij}}{\mathcal P}_{\bf k}^{j,m}{\mathcal P}_{\bf k}^{i,m*}-F_{ij}|{\mathcal P}_{\bf k}^{j,m}|^2\Big)\Big]\nonumber.
\end{eqnarray}
The summation $\sum_{j\in\langle i,j\rangle}$ follows the same rule as that in Eq. (\ref{gamma}). And the temperature-dependent parameters $\eta_{ij}$ in Eq. (\ref{MF}) are expressed as
\begin{eqnarray}\label{eta}
\eta_{ij}=\frac1{N}\sum_{\bf p}\sum_{l=1}^{20}&&(n_{l\bf p}+\delta_{i}^{l})\\
&&\times\Big(|{\mathcal P}_{\bf p}^{i,l}|^2-F_{ij}e^{-i{\bf p}\cdot{\bf r}_{ij}}{\mathcal P}_{\bf p}^{j,l*}{\mathcal P}_{\bf p}^{i,l}\Big),\nonumber
\end{eqnarray}
with
\begin{equation}\begin{aligned}
&J_{ij}=J_{aa}\quad \mbox{and}\quad F_{ij}=1, &&\quad\, \mbox{when}\;i,j\in\{1,\cdots,8\},\nonumber\\
&J_{ij}=J_{dd}\quad \mbox{and}\quad F_{ij}=1, &&\quad\, \mbox{when}\;i,j\in\{9,\cdots,20\},\nonumber\\
&J_{ij}=J_{ad}\quad \mbox{and}\quad F_{ij}=-1, &&\quad \text{ otherwise}.\nonumber
\end{aligned}\end{equation} 
After implementing the mean-field approximation, it is apparent that the magnon eigenstates $\alpha_{1\sim8}$ and $\beta_{1\sim12}$ undergo a renormalization. In instances where the following conditions are met, $\delta_{i}^{l}=1$ :  (1) $i\in\{1,\cdots,8\}$ and $l\in\{9,\cdots,20\}$; (2) $i\in\{9,\cdots,20\}$ and $l\in\{1,\cdots,8\}$. Otherwise, $\delta_{i}^{l}$ evaluates to zero. 

Taking the correction $\Delta\tilde\varepsilon_{m,k}$ into consideration, the temperature-dependent magnon energy spectrum of YIG is obtained as,
\begin{equation}
\tilde\varepsilon_{m,k}=\varepsilon_{m,{\bf k}}+\Delta\tilde\varepsilon_{m,{\bf k}},
\end{equation}
where $\varepsilon_{m,k}$ are determined by the diagonalization of Eq. (\ref{HH2}), referring to the magnon energies without interactions.
When $m\in\{1,2,\cdots,8\}$, $\tilde \varepsilon_{1\sim8,{\bf k}}$ signify the energies of the $\alpha_{1\sim8}$ bands, which are denoted as  $\tilde \varepsilon_{1\sim8,{\bf k}}^\alpha$. For $m\in\{9,10,\cdots,20\}$, $\tilde \varepsilon_{9\sim20,{\bf k}}$ represent the energies of the $\beta_{1\sim12}$ bands, denoted as $\tilde \varepsilon_{1\sim12,{\bf k}}^\beta$.
Thus, the self-consistent procedures are established. 

\subsection{The current and current-current correlation functions}

\begin{figure*}[hbtp]
\centering
\includegraphics[width=\textwidth]{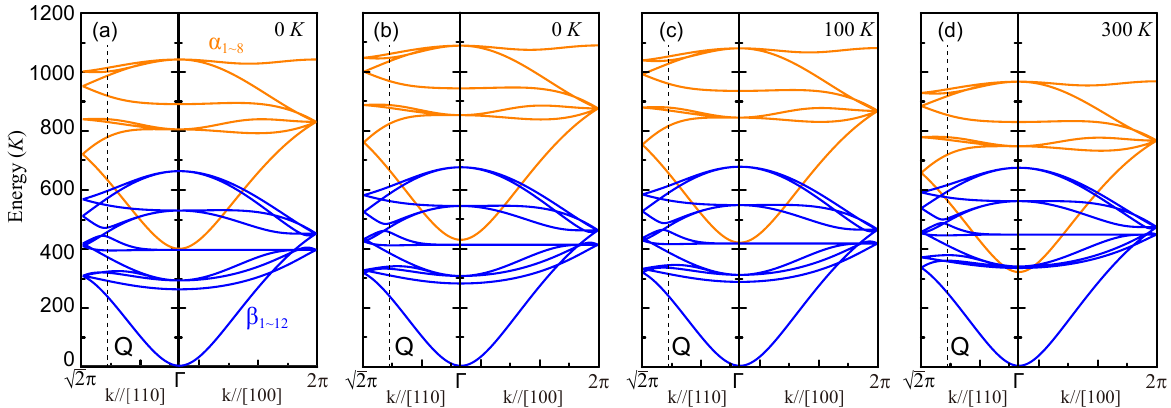}
\caption{(a), The magnon spectra of YIG obtained directly from Eq. (\ref{HH2}), with the magnon-magnon interactions not included. (b-d), The magnon spectra of YIG, inclusive of magnon-magnon interactions, are depicted along the $[110]$ and $[100]$ directions at temperatures of $T=0K$, $T=100K$, and $T=300K$ respectively. The orange bands, from low to high, are denoted as $\alpha_1,\alpha_2,\cdots,\alpha_8$, while the blue bands, similarly ordered from low to high, are represented as $\beta_1,\beta_2,\cdots,\beta_{12}$. The dashed lines indicate the point $Q$ at which no degeneracy occurs.}\label{YIG-e}
\end{figure*}

In this subsection, we'd like to discuss the temperature-dependent behaviors of spin transport coefficients, namely the spin Seebeck coefficient (${S_m}$), spin conductivity (${\sigma_m}$), and thermal conductivity (${\kappa_m}$). In a magnetic system, these coefficients are intricately linked to the interaction between spin current and thermal current \cite{DanielLoss2015,mahan2013many}, i.e. ${S}_m={L}_{12}/L_{11}$, ${\sigma}_m={L}_{11}$, and ${\kappa}_m=(L_{22}-{L_{12}L_{21}}/{L_{11}})/T$. These response functions, denoted as ${L}_{\mu\nu}$, are defined by
\begin{equation}\label{JS-JQ}
\begin{pmatrix}\bm{J_S}\\ \bm{J_Q}\end{pmatrix}=\begin{pmatrix}{L}_{11} & {L}_{12}\\ {L}_{21} & {L}_{22}\end{pmatrix}\begin{pmatrix}\bm{\nabla B}\\-{\bm{\nabla} T}/T\end{pmatrix}
\end{equation}
where $\bm{J_S}$ is the spin current, $\bm{J_Q}$ is the thermal current, $\bm{\nabla} T$ is the temperature gradient, and $\bm{\nabla B}$ is the magnetic-field gradient \cite{DanielLoss2015,Arakawa2022}. The selection of current and force in Eq. (\ref{JS-JQ}) indicates that Onsager's relation is valid: ${L_{12}}={L_{21}}$ \cite{Onsager1931,Casimir1945,Kubo1957,Oji1985}.
Moreover, there's no chemical potential for spin systems, so $\bm{J_Q}=\bm{J_E}$ \cite{mahan2013many}, with $\bm{J_E}$ the energy current carried by magnons.
${L}_{\mu\nu}$'s are related directly to the corresponding retarded current-current correlation functions $\Phi_{\mu\nu}^R(\omega)$ \cite{eliashberg1962transport,Luttinger1964,Kubo1957, mahan2013many,Fukuyama1970}, as shown in Appendix~\ref{app: C}.
Substituting the spin current and energy current into these current-current correlation functions, the expression of the response functions is obtained as \cite{Arakawa2018,Fukuyama1970},
\begin{equation}\label{Lmunu}\begin{aligned}
L_{\mu\nu}=\frac{1}{N}\sum_{m,l,{\bf k}}\!\!\Lambda_{\mu\nu}^{ml}({\bf k})\int_{-\infty}^\infty\frac{d\omega}{4\pi}\frac{\partial n(\omega)}{\partial\omega}A_m({\bf k},\omega)A_l ({\bf k},\omega),
\end{aligned}\end{equation}
where $A_m ({\bf k},\omega)=-2{\rm Im}G^{\rm R}_m({\bf k},\omega)$ is the spectral function of magnon, and $n(\omega)$ is the Bose distribution function. The matrix element $\Lambda_{\mu\nu}^{ml}(\bf k)$ of ${\bm\Lambda_{\mu\nu}}(\bf k)$ serves as the vertex function for the current-current correlation functions $\Phi_{\mu\nu}(i\Omega_n)$. 

Specifically, $\bm\Lambda_{\mu\nu}(\bf k)$ is defined as 
\begin{equation}\begin{aligned}
&{\bm\Lambda_{11}}({\bf k})={\bm{\mathcal P}}_{\bf k}^\dag\frac{-\partial {\bm{\mathcal H}}_{\bf k}}{\partial {\bf k}}{\bm{\mathcal P}}_{\bf k}{\bm{\mathcal P}}_{\bf k}^\dag\frac{-\partial {\bm{\mathcal H}}_{\bf k}}{\partial {\bf k}}{\bm{\mathcal P}}_{\bf k},\\
&{\bm\Lambda_{12}}({\bf k})={\bm{\mathcal P}}_{\bf k}^\dag\frac{-\partial {\bm{\mathcal H}}_{\bf k}}{\partial {\bf k}}{\bm{\mathcal P}}_{\bf k}{\bm{\mathcal P}}_{\bf k}^\dag{\frac12}\frac{\partial [{\bm{\mathcal H}_{\bf k}{\bf g}\bm{\mathcal H}}_{\bf k}]}{\partial {\bf k}}{\bm{\mathcal P}}_{\bf k},\\
&{\bm\Lambda_{22}}({\bf k})={\bm{\mathcal P}}_{\bf k}^\dag{\frac12}\frac{\partial [{\bm{\mathcal H}_{\bf k}{\bf g}\bm{\mathcal H}}_{\bf k}]}{\partial {\bf k}}{\bm{\mathcal P}}_{\bf k}{\bm{\mathcal P}}_{\bf k}^\dag{\frac12}\frac{\partial [{\bm{\mathcal H}_{\bf k}{\bf g}\bm{\mathcal H}}_{\bf k}]}{\partial {\bf k}}{\bm{\mathcal P}}_{\bf k}.
\end{aligned}\end{equation}
The imaginary parts of retarded magnon Green's functions are given by :
\begin{equation}\begin{aligned}
{\rm Im} G^{\rm R}_{m}({\bf k},\omega)=\frac{-\gamma}{(\omega-\tilde\varepsilon_{m,{\bf k}})^2+\gamma^2}&, \quad m=1\cdots8;\\
{\rm Im} G^{\rm R}_{m}({\bf k},\omega)=\frac{\gamma}{(\omega+\tilde\varepsilon_{m,{\bf k}})^2+\gamma^2}&, \quad m=9\cdots20.
\end{aligned}\end{equation}
Here, $\gamma$ is a temperature-dependent parameter that describes the scattering rate of magnons in YIG. When only considering the magnon-magnon interaction of the four-magnon process, $\gamma$ can be given as $\gamma=\gamma_0+\gamma_1T+\gamma_2T^2$, where $\gamma_0=0.4$, $\gamma_1=10^{-4}$, and $\gamma_2=2.5\times10^{-5}$ \cite{Arakawa2022}.

\section{Results and Discussion}

\subsection{Temperature dependence of magnon spectra of YIG}

Taking the self-consistent mean-field correction for the magnon-magnon interactions into consideration, the temperature-dependent magnon energy spectra of YIG are obtained, as depicted in Figs. \ref{YIG-e}(b), \ref{YIG-e}(c) and \ref{YIG-e}(d). Where, Fig. \ref{YIG-e}(a) illustrates the bare magnon spectrum, excluding the magnon-magnon interactions. This bare spectrum is attained through the direct diagonalization of Eq. (\ref{HH2}).

From Eqs. (\ref{addij}--\ref{eta}), we find that in addition to the correction that varies with temperature, the magnon-magnon interaction also brings a correction to magnon energies at zero temperature (the term $\delta_i^l$ in Eq. (\ref{eta}) ).
Initially, we incorporated the temperature-invariant correction, resulting in the adjusted magnon energy spectrum depicted in Fig. \ref{YIG-e}(b).
Compared with Fig. \ref{YIG-e}(a), the spectrum in Fig. \ref{YIG-e}(b) shows a noticeable increase, indicating that the magnon-magnon interactions contribute to an enhancement in the magnon energy at zero temperature.
This zero-temperature enhancement has been known for a long time, first calculated by Anderson and Kubo \cite{Anderson1952,Kubo1952,Oguchi1960}, and verified in antiferromagnetic materials such as La$_2$CuO$_4$ \cite{Igarashi1992, Canali1992}.
Subsequently, the temperature-dependent corrections stemming from the magnon-magnon interactions are taken into account and the magnon spectra of YIG at $100K$ and $300K$ are depicted in Fig. \ref{YIG-e}(c) and \ref{YIG-e}(d), respectively.
The eight orange $\alpha$ bands of higher energy decrease with increasing temperature, while the twelve blue $\beta$ bands of lower energy show little change with temperature.

\begin{figure}[b]
\centering
\includegraphics[width=0.45\textwidth]{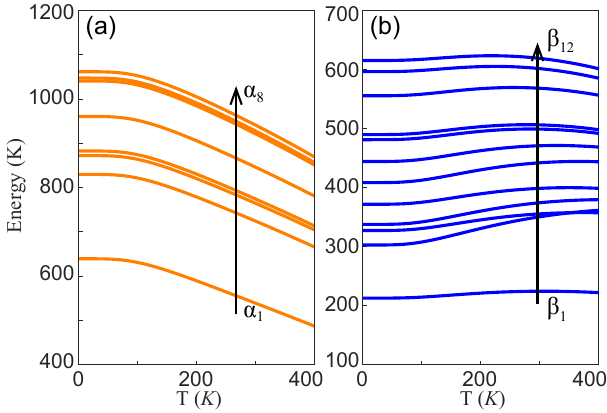}
\caption{The temperature dependence of magnon spectra of YIG at point $Q$ depicted in Fig. \ref{YIG-e}. (a) The temperature dependence of $\alpha_{1\sim8}$ bands. (b) The temperature dependence of $\beta_{1\sim12}$ bands.}\label{YIG-e-2}
\end{figure}

\begin{figure*}[htpb]
\centering
\includegraphics[width=1.0\textwidth]{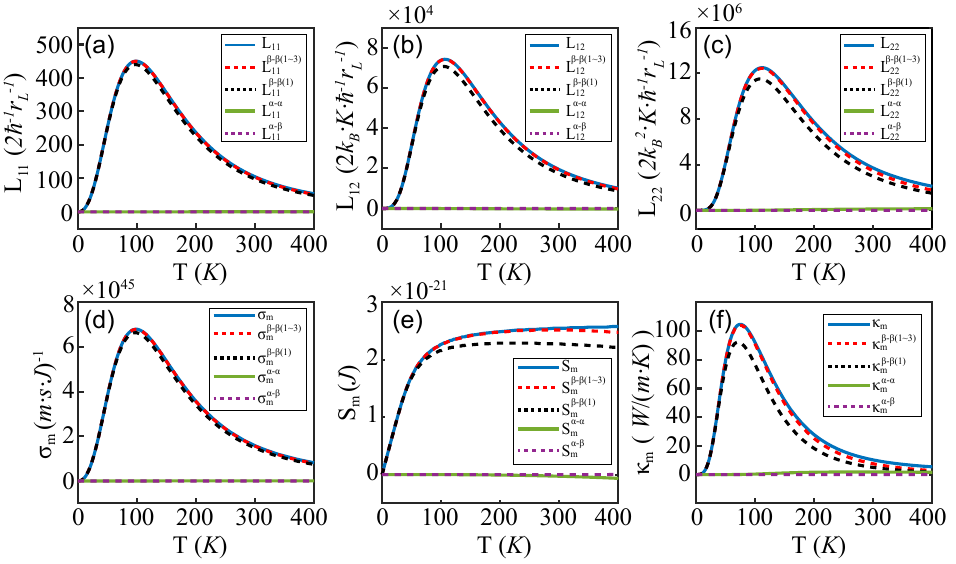}
\caption{The variation of transport coefficients $L_{\mu\nu}$'s with temperature: (a) $L_{11}$, (b) $L_{12}$, (c) $L_{22}$. The blue solid line, the green solid line, and the violet dashed line represent the quantities shown in Eq. (\ref{Lmunu-2}). The red dashed line signifies the contributions from $\beta_1,\beta_{2}$, and $\beta_{3}$ magnon modes.The black dashed line signifies the contributions from only $\beta_1$ mode. The corresponding curves of $\sigma_{m}$, $S_{m}$, and $\kappa_{m}$ with temperature are shown in (d), (e), and (f) respectively. Where, $r_L=12.56\AA$ is the lattice constant of conventional cell of YIG, $W$ denotes the unit watt, and $J$ denotes the unit joule.}\label{YIG-L-123}
\end{figure*}

To see this clearly, we have selected a specific $Q$ point and plotted the temperature-dependent variation of the 20 magnon energies at this $Q$ point, as shown in Fig. \ref{YIG-e-2}. The $Q$ point lies in the $[110]$ direction, shown by the dashed line in Fig. \ref{YIG-e}, where the bands are non-degenerate. Fig. \ref{YIG-e-2}(a) shows the temperature dependences of $\alpha_{1\sim8}$ bands at the $Q$ point. The results demonstrate a consistent decrease in the energy of the $\alpha$ bands. Fig. \ref{YIG-e-2}(b) displays the temperature dependences of $\beta_{1\sim12}$ bands at the $Q$ point. It can be seen that within the temperature range studied, the $\beta$ bands exhibit minimal variation with temperature.

These results are consistent with the earlier INS experiments \cite{Plant1977}. 
In the INS experiments, the spin stiffness of acoustic branch was observed to change little with temperature, which contrary to common sense.
In the common sense the spin stiffness should be softened with the increase of temperature \cite{Rezende2020fundamentals,Barker2016}.
Moreover, the steep optic branch (namely $\alpha_1$ band in our notations) measured in the INS experiments falls to lower energies with increasing temperature \cite{Plant1977}.
At $83K$, the energy of steep optic branch at point $\Gamma$ is approximately 8.24 THz, while at $295K$, it is about 6.4 THz.
Our results are in agreement with the experiments, with $\tilde\varepsilon_{1,\bm k}^\alpha\approx$ 8.7 THz at $100K$ and decreasing to 6.7 THz at $300K$.

These agreements lead us to believe that the minimal temperature-induced variation in the $\beta$ bands is primarily driven by the interband magnon-magnon interactions. Our results also exhibit good agreement with theoretical calculations, including those derived from atomistic spin simulations \cite{Barker2016} and self-consistent mean-field calculations \cite{Shen2018}.

\subsection{Temperature dependence of $\sigma_m, S_m$ and $\kappa_m$}

Here, we study the temperature dependence of these three transport coefficients: $\sigma_m, S_m$, and $\kappa_m$.
We analyze the 20 magnon bands of YIG and identify which specific bands contribute predominantly to these three transport coefficients.
From Eq.(\ref{Lmunu}), it is evident that $L_{\mu\nu}$'s are determined by the products of the magnon spectrum functions.
In addition to self-multiplication within a band, there is also multiplication between the magnon spectrum functions of different bands.
Therefore, it is important to investigate and understand the influence of the coupling between different magnon bands on the magnon transport coefficients.

We divide the total contribution into three parts:
\begin{equation}\label{Lmunu-2}
L_{\mu\nu}=L_{\mu\nu}^{\alpha-\alpha}+L_{\mu\nu}^{\alpha-\beta}+L_{\mu\nu}^{\beta-\beta},
\end{equation}
where the differences between $L_{\mu\nu}^{\alpha-\alpha}$, $L_{\mu\nu}^{\alpha-\beta}$, and $L_{\mu\nu}^{\beta-\beta}$ lie solely in the summation over the indices $m$ and $l$, as indicated in Eq. (\ref{Lmunu}). These indices range as follows:
\begin{equation*}\begin{aligned}
&\mbox{for}\; L_{\mu\nu}^{\alpha-\alpha}, m,l\in\{1,2,\cdots,8\};\\
&\mbox{for}\; L_{\mu\nu}^{\beta-\beta}, m,l\in\{9,10,\cdots,20\};\\
&\mbox{for}\; L_{\mu\nu}^{\alpha-\beta}, \mbox{otherwise.}
\end{aligned}\end{equation*}
However, this division in Eq. (\ref{Lmunu-2}) is inadequate because the contribution to $L_{\mu\nu}$ mainly comes from $L_{\mu\nu}^{\beta-\beta}$, or even more specifically, from the three lowest-energy $\beta$ bands denoted as $\beta_1, \beta_2, \beta_3$. As depicted in Fig. \ref{YIG-L-123}, the curve for $L_{\mu\nu}$ is closely aligns with that for $L_{\mu\nu}^{\beta-\beta(1\sim3)}$, where the superscript of $L_{\mu\nu}^{\beta-\beta(1\sim3)}$ indicating that the contributions stem from $\beta_1, \beta_2$, and $\beta_3$ bands.

We plot the variation of each $L_{\mu\nu}$ component with temperature in Figs. \ref{YIG-L-123}(a), \ref{YIG-L-123}(b) and \ref{YIG-L-123}(c). The contributions of $L_{\mu\nu}^{\alpha-\alpha}$ and $L_{\mu\nu}^{\alpha-\beta}$ are minimal, which aligns well with physical reality. This is because the eight $\alpha$ bands have excessively high energy levels, making it virtually impossible for magnons to be excited to these bands within the depicted temperature range. Furthermore, the curve representing $L_{\mu\nu}^{\beta-\beta(1\sim3)}$ closely aligns with the $L_{\mu\nu}$ curve, suggesting that the higher-energy bands, including $\beta_4,\beta_5,\cdots, \beta_{12}$, have a negligible effect  on $L_{\mu\nu}$. As shown in Figs. \ref{YIG-L-123}(b) and \ref{YIG-L-123}(c), the black dashed curve representing the contribution from the $\beta_1$ band deviates from the $L_{\mu\nu}$ curve. Thus considering only the contribution of the lowest-energy $\beta_1$ band seems insufficient.

Subsequently, we present the curves of $\sigma_m,S_m,\kappa_m$ as a function of temperature, as shown in Fig. \ref{YIG-L-123}(d), \ref{YIG-L-123}(e), \ref{YIG-L-123}(f). Since $\sigma_m=L_{11}$, the curve of the magnon conductivity is identical to that of $L_{11}$. Fig. \ref{YIG-L-123}(d) shows a small but noticeable deviation between $\sigma_m^{\beta-\beta(1)}$ and $\sigma_m$, whereas $\sigma_m^{\beta-\beta(1\sim3)}$ almost overlaps with $\sigma_m$. This suggests that in YIG, the transport of the $\beta_1, \beta_{2},\beta_{3}$ magnons collectively accounts for nearly all of the magnon conductivity. Notably, focusing on the contribution of the $\beta_1$ magnon alone still provides satisfactory results.
However, the situation is different for $S_m$ and $\kappa_m$. Figs. \ref{YIG-L-123}(e) and \ref{YIG-L-123}(f) show that the curve for $S_m^{\beta-\beta(1)}$ ($\kappa_m^{\beta-\beta(1)}$) exhibits a significant deviation from that of $S_m$ ($\kappa_m$). In contrast, the curves stemming from $\beta_1,\beta_2$ and $\beta_3$ bands aligns more closely with the curves of $S_m$ and $\kappa_m$ that incorporates all contributions.

Based on the above discussions, we conclude that considering the contribution of the $\beta_1$ magnon alone can yield satisfactory results for the magnon conductivity $\sigma_m$ of YIG. However, for the spin Seebeck coefficient $S_m$ and magnetic thermal conductivity $\kappa_m$, it is necessary to include the contributions of higher-energy bands such as $\beta_{2}$ and $\beta_{3}$.

\begin{figure}[htpb]
\centering
\includegraphics[width=0.45\textwidth]{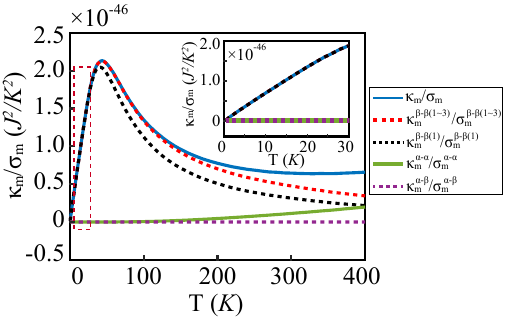}
\caption{The curves of the ratio $\kappa_m/\sigma_m$ as a function of temperature. The values of $\sigma_m$ and $\kappa_m$ are from Fig. \ref{YIG-L-123}(d) and Fig. \ref{YIG-L-123}(f), respectively. The blue solid line, green solid line, and purple dashed line represent the contributions of scattering between all magnon modes, scattering between $\alpha_{1-8}$ modes, and scattering between $\alpha_{1-8}$ modes and $\beta_{1-12}$ modes to $\kappa_m/\sigma_m$, respectively. The red dashed line and black dashed line represent the contributions of scattering between $\beta_{1-3}$ modes and scattering in $\beta_{1}$ mode to $\kappa_m/\sigma_m$, respectively. The inset shows this ratio as a function of temperature in the range of $0K\sim30K$. It can be seen that $\kappa_m/\sigma_m$ is predominantly influenced by the acoustic magnon at low temperatures, exhibiting a linear dependence on temperature.}\label{YIG-WFlaw}
\end{figure}

Then we checked whether our transport coefficients satisfy the Wiedemann-Franz law for magnon transport \cite{DanielLoss2015,Nakata2017}, which is an analog to the charge Wiedemann-Franz law.
For charge transport, the Wiedemann-Franz law states that the ratio of thermal conductivity to electrical conductivity of electrons at low temperatures is linearly related to temperature \cite{Franz1853,Kittel2004}.
Here, we calculate the ratio of $\kappa_m/\sigma_m$ and plot its variation with temperature in Fig. \ref{YIG-WFlaw}. 
We observe a linear temperature dependence of $\kappa_m/\sigma_m$ within the lower temperature range of $0\sim30K$.
This linear dependence is consistent with the theoretical results in Ref. \cite{DanielLoss2015,Nakata2017}.
Furthermore, at low temperatures, solely the acoustic magnons ($\beta_1$ mode) are excited, hence it is exclusively the $\beta_1$ magnons that contribute to the Wiedemann-Franz law for magnon transport.
This is clearly shown in the inset of Fig. \ref{YIG-WFlaw}.

\section{Conclusion}

In this paper, we have conducted a study of the magnon spectra and the spin Seebeck effect in YIG.
After the HP transformation and Bogoliubov transformation, the energy spectra without magnon-magnon interaction and the para-unitary matrix are obtained.
Then, the para-unitary transformation and the mean-field approximation are performed on the magnon-magnon interaction terms, resulting in the temperature-dependent magnon spectra of YIG.
Using these spectra, we analyzed the temperature-induced variations in magnon conductivity ($\sigma_m$), spin Seebeck coefficient ($S_m$), and magnon thermal conductivity ($\kappa_m$) in YIG.

In our notation, $\alpha_{1\sim8}$ represent the eight higher-energy bands of YIG, while $\beta_{1\sim12}$ represent the 12 lower-energy bands, with $\beta_{1}$, $\beta_{2}$, and $\beta_{3}$ specifically comprising the three bands with the lowest energies.
The calculation results of the magnon spectra in YIG reveal that as the temperature increases, the magnon-magnon interaction prompts noticeable declines in the $\alpha$ bands, whereas the $\beta$ bands exhibit little variations with temperature, as shown in Fig. \ref{YIG-e} and Fig. \ref{YIG-e-2}.
Moreover, the energy of $\alpha_1$ band at the $\Gamma$ point is about 8.7 THz at $100K$ and decreases to 6.7 THz at $300K$.
These results agree well with the observation of earlier INS experiments, as well as the results of the theoretical calculations in recent years.
In addition, the calculation results of the spin Seebeck effect indicate that when solely focusing on the magnon conductivity of YIG, incorporating the contribution of the $\beta_1$ band is adequate, while the inclusion of the contributions from the $\beta_{2}$ and $\beta_{3}$ bands does not yield a notable difference, as shown in Fig \ref{YIG-L-123}(d).
To accurately investigate the spin Seebeck coefficient and magnon thermal conductivity, it is imperative to consider the collective contributions of the three lowest energy bands: $\beta_1$,$\beta_{2}$,$\beta_{3}$, as shown in Fig. \ref{YIG-L-123}(e), (f).
The results calculated using only $\beta_1$ show a noticeable deviation over a large energy range compared to the full band calculations. Finally, we find our results satisfy the Wiedemann-Franz law for magnon transport. The major contribution to the Weidmann-Franz law comes from the $\beta_1$ mode. 

\section*{Acknowledgement}
The authors will thank Prof. Huaisong Zhao and Prof Yongping Du for the constructive suggestion, and thank Dr. Di Wang, Dr. Xinhai Tu, Dr. Songsong Yan, Dr. Xingchuan Zhu, and Dr. Minghuan Zeng for the helpful discussion. This work was supported by the NSFC (No. 12188101, 12147139, 12404121), the National Key R\&D Program of China (Grant No. 2022YFA1403601), the Innovation Program for Quantum Science and Technology (No. 2021ZD0301902, 2024ZD0300101), and the Natural Science Foundation of Jiangsu Province (No. BK20243011).

\begin{widetext}
\appendix
\section{\MakeUppercase{The quadratic and interaction Hamiltonian}}
\label{app: A}
\renewcommand{\theequation}{A\arabic{equation}}  
\setcounter{equation}{0} 
After the HP transformation, the Hamiltonian of YIG in Eq. (\ref{HHH}) can be expressed as $H=H^{(0)}+H^{(2)}+H^{(4)}+\cdots,$ where the quadratic Hamiltonian $H^{(2)}$ can be divided as $H^{(2)}=H_{aa}^{(2)}+H_{dd}^{(2)}+H_{ad}^{(2)}+H_{B}^{(2)}$. Each part of $H^{(2)}$ is shown as follows: 
\begin{equation}\begin{aligned}
H^{(2)}_{aa}\!=&J_{aa}S\sum_{n=1}^N \sum_{i=1}^8 \Big(16\;a_{\mathbf{R}_{n,i}}^\dagger a_{\mathbf{R}_{n,i}}\!-2\!\!\!\sum_{\mathbf{|r_{ij}|}=\rm{r_{aa}}}\!\!\!a_{\mathbf{R}_{n,i}}^\dagger a_{\mathbf{R}_{n,i}\!+\!\mathbf{r}_{ij}}\Big),
H^{(2)}_{dd}\!=J_{dd}S\sum_{n=1}^N \sum_{i=1}^{20} \Big(8\;d_{\mathbf{R}_{n,i}}^\dagger d_{\mathbf{R}_{n,i}}\!-2\!\!\!\sum_{\mathbf{|r_{ij}|}=\rm{r_{dd}}}\!\!\!d_{\mathbf{R}_{n,i}}^\dagger d_{\mathbf{R}_{n,i}\!+\!\mathbf{r}_{ij}}\Big),\\
H^{(2)}_{ad}\!=&J_{ad}S\sum_{n=1}^N \sum_{i=1}^{8} \Big(-12\;a_{\mathbf{R}_{n,i}}^\dagger a_{\mathbf{R}_{n,i}}-2\!\!\!\sum_{\mathbf{|r_{ij}|}=\rm{r_{ad}}}\!\!\!a_{\mathbf{R}_{n,i}}^\dagger d_{\mathbf{R}_{n,i}\!+\!\mathbf{r}_{ij}}^\dagger \Big)+J_{ad}S\sum_{n=1}^N \sum_{i=9}^{20} \Big(\!-8\;d_{\mathbf{R}_{n,i}}^\dagger d_{\mathbf{R}_{n,i}}\!\!-2\!\!\!\sum_{\mathbf{|r_{ij}|}=\rm{r_{ad}}}\!\!\!d_{\mathbf{R}_{n,i}}a_{\mathbf{R}_{n,i}\!+\!\mathbf{r}_{ij}}\Big),\\
H_{B}^{(2)}=&-\sum_{n=1}^N \sum_{i=1}^8g\mu_B B a_{\mathbf{R}_{n,i}}^\dagger a_{\mathbf{R}_{n,i}}+\sum_{n=1}^N \sum_{i=9}^{20}g\mu_BBd_{\mathbf{R}_{n,i}}^\dagger d_{\mathbf{R}_{n,i}}.
\end{aligned}\end{equation}
Moreover, after the HP transformation, the interaction Hamiltonian $H^{(4)}$ can be expressed into three parts, $H^{(4)}=H_{aa}^{(4)}+H_{dd}^{(4)}+H_{ad}^{(4)}$. Each part can be expressed as:
\begin{equation}\begin{aligned}
H_{aa}^{(4)}=&\frac12J_{aa}\sum_{n=1}^N \sum_{i=1}^8\!\sum_{\mathbf{|r_{ij}|}=\rm{r_{aa}}}
\Big(a_{\mathbf{R}_{n,i}}^\dagger a_{\mathbf{R}_{n,i}\!+\!\mathbf{r}_{ij}}^\dagger a_{\mathbf{R}_{n,i}\!+\!\mathbf{r}_{ij}} a_{\mathbf{R}_{n,i}\!+\!\mathbf{r}_{ij}}+a_{\mathbf{R}_{n,i}}^\dagger a_{\mathbf{R}_{n,i}}^\dagger a_{\mathbf{R}_{n,i}}a_{\mathbf{R}_{n,i}\!+\!\mathbf{r}_{ij}}-2a_{\mathbf{R}_{n,i}}^\dagger a_{\mathbf{R}_{n,i}}a_{\mathbf{R}_{n,i}\!+\!\mathbf{r}_{ij}}^\dagger a_{\mathbf{R}_{n,i}\!+\!\mathbf{r}_{ij}}\Big)\\
H_{dd}^{(4)}=&\frac12J_{dd}\sum_{n=1}^N \sum_{i=9}^{20}\!\sum_{\mathbf{|r_{ij}|}=\rm{r_{dd}}}
\Big( d_{\mathbf{R}_{n,i}}^\dagger d_{\mathbf{R}_{n,i}} d_{\mathbf{R}_{n,i}}d_{\mathbf{R}_{n,i}\!+\!\mathbf{r}_{ij}}^\dagger+d_{\mathbf{R}_{n,i}}d_{\mathbf{R}_{n,i}\!+\!\mathbf{r}_{ij}}^\dagger d_{\mathbf{R}_{n,i}\!+\!\mathbf{r}_{ij}}^\dagger d_{\mathbf{R}_{n,i}\!+\!\mathbf{r}_{ij}}-2d_{\mathbf{R}_{n,i}}^\dagger d_{\mathbf{R}_{n,i}}d_{\mathbf{R}_{n,i}\!+\!\mathbf{r}_{ij}}^\dagger d_{\mathbf{R}_{n,i}\!+\!\mathbf{r}_{ij}}\Big)\\
H_{ad}^{(4)}
=&\frac12J_{ad}\sum_{n=1}^N \sum_{i=1}^8\!\sum_{\mathbf{|r_{ij}|}=\rm{r_{ad}}}
\Big(\;a_{\mathbf{R}_{n,i}}^\dagger d_{\mathbf{R}_{n,i}\!+\!\mathbf{r}_{ij}}^\dagger d_{\mathbf{R}_{n,i}\!+\!\mathbf{r}_{ij}}^\dagger d_{\mathbf{R}_{n,i}\!+\!\mathbf{r}_{ij}}+a_{\mathbf{R}_{n,i}}^\dagger a_{\mathbf{R}_{n,i}}^\dagger a_{\mathbf{R}_{n,i}}d_{\mathbf{R}_{n,i}\!+\!\mathbf{r}_{ij}}^\dagger\Big)\\
+&\frac12J_{ad}\sum_{n=1}^N \sum_{i=9}^{20}\!\sum_{\mathbf{|r_{ij}|}=\rm{r_{ad}}}
\Big(d_{\mathbf{R}_{n,i}}a_{\mathbf{R}_{n,i}\!+\!\mathbf{r}_{ij}}^\dagger a_{\mathbf{R}_{n,i}\!+\!\mathbf{r}_{ij}} a_{\mathbf{R}_{n,i}\!+\!\mathbf{r}_{ij}}+d_{\mathbf{R}_{n,i}}^\dagger d_{\mathbf{R}_{n,i}}d_{\mathbf{R}_{n,i}}a_{\mathbf{R}_{n,i}\!+\!\mathbf{r}_{ij}}+4d_{\mathbf{R}_{n,i}}^\dagger d_{\mathbf{R}_{n,i}}a_{\mathbf{R}_{n,i}\!+\!\mathbf{r}_{ij}}^\dagger a_{\mathbf{R}_{n,i}\!+\!\mathbf{r}_{ij}}\Big)
\end{aligned}\end{equation}
Then the Fourier transformation $a_{\mathbf{R}_{n,i}}=\frac1{\sqrt{N}}\sum_{\bf k}e^{i\bf k\cdot \bf{R}_{n,i}}a_{i\bf k}$ and $d_{\mathbf{R}_{n,i}}=\frac1{\sqrt{N}}\sum_{\bf k}e^{i\bf k\cdot \bf{R}_{n,i}}d_{i\bf k}$ are performed on $H^{(2)}$ and $H^{(4)}$. These yield:
\begin{equation}\begin{aligned}
&H^{(2)}_{aa}\!=J_{aa}S\sum_{\bf k} \sum_{i=1}^8 \Big(16\;a_{i{\bf k}}^\dagger a_{i{\bf k}}\!-2\!\sum_{j\in\langle i,j\rangle}^{a\to a}e^{i\bf{k}\cdot(\bm{r}_j-\bf{r}_i)}a_{i{\bf k}}^\dagger a_{j{\bf k}}\Big),H^{(2)}_{dd}\!=J_{dd}S\sum_{\bf k} \sum_{i=9}^{20} \Big(8\;d_{i-{\bf k}}^\dagger d_{i-{\bf k}}\!-2\!\sum_{j\in\langle i,j\rangle}^{d\to d}e^{i\bf{k}\cdot(\bm{r}_j-\bf{r}_i)}d_{i-{\bf k}}^\dagger d_{j-{\bf k}}\Big),\\
&H^{(2)}_{ad}\!=J_{ad}S\sum_{\bf k}\sum_{i=1}^8 \Big(-12\;a_{i{\bf k}}^\dagger a_{i{\bf k}}\!-2\!\sum_{j\in\langle i,j\rangle}^{a\to d}e^{i\bf{k}\cdot(\bm{r}_j-\bf{r}_i)}a_{i{\bf k}}^\dagger d_{j-{\bf k}}^{\dag}\Big)\;+\;J_{ad}S\sum_{\bf k}\sum_{i=9}^{20} \Big(-8\;d_{i-{\bf k}}^\dagger d_{i-{\bf k}}\!-2\!\sum_{j\in\langle i,j\rangle}^{d\to a}e^{i\bf{k}\cdot(\bm{r}_j-\bf{r}_i)}d_{i-{\bf k}} a_{j{\bf k}}\Big),\\
&H_{B}^{(2)}=-\sum_{\bf k} \sum_{i=1}^8g\mu_B B a_{i{\bf k}}^\dagger a_{i{\bf k}}+\sum_{\bf k}\sum_{i=9}^{20}g\mu_BBd_{i-{\bf k}}^\dagger d_{i-{\bf k}}.
\end{aligned}\end{equation}
Here, $\sum_{j\in\langle i,j\rangle}^{a\to a}$  is a summation over the $j$th site, which are restricted to the neighbors of the $i$th site, connected by the $J_{aa}$ bond. 
The superscript $a\rightarrow a$ indicates that the ions on $i$ and $j$ sites are both $a$-type iron ions. $\sum_{j\in\langle i,j\rangle}^{a\to d}$, $\sum_{j\in\langle i,j\rangle}^{d\to a}$ and $\sum_{j\in\langle i,j\rangle}^{d\to d}$ follow the similar habits as that of $\sum_{j\in\langle i,j\rangle}^{a\to a}$.
For example, \(\sum_{j\in\langle i,j\rangle}^{a\to d}\) sums over \(j\) in the \(d\) sites when \(i\) belongs to the \(a\) sites, and the sites $i$ and $j$ are connected by the $J_{ad}$ bond.
Moreover, for the four types of summation notations, ${{\bf r}_{ij}}={{\bf r}_{j}}-{{\bf r}_{i}}$ satisfies $|{{\bf r}_{ij}}|={\rm r_{aa}}$, ${\rm r_{ad}}$, ${\rm r_{ad}}$ and ${\rm r_{ad}}$, respectively. And we can summarize the above expressions of $H^{(2)}$ into the matrix formats as depicted in Eq. (\ref{HH2}) in the main text. The interaction terms with four magnons are obtained as follows: 
\begin{equation}\label{HH4app}\begin{aligned}
H_{aa}^{(4)}=&\frac{J_{aa}}{N}\sum_{{\bf k}_1-{\bf k}_4}\delta_{{\bf k}_1+{\bf k}_2-{\bf k}_3-{\bf k}_4}\!\!\sum_{i=1}^8\sum_{j\in\langle i,j\rangle}^{a\to a}\Big(e^{i\bf{k}_1(\bm{r}_j-\bf{r}_i)}a_{i{\bf k}_4}^\dagger a_{i{\bf k}_3}^\dagger a_{i{\bf k}_2}\;a_{j{\bf k}_1}+e^{i\bf{k}_4(\bm{r}_j-\bf{r}_i)}a_{i{\bf k}_4}^\dagger a_{j{\bf k}_3}^\dagger a_{j{\bf k}_2}a_{j{\bf k}_1}\\
&\hspace{28em}-2e^{i(\bf{k}_1-\bf{k}_4)\cdot(\bm{r}_j-\bf{r}_i)}a_{j{\bf k}_4}^\dagger a_{i{\bf k}_3}^\dagger a_{i{\bf k}_2} a_{j{\bf k}_1}\Big)\\
H_{dd}^{(4)}=&\frac{J_{dd}}{2N}\sum_{{\bf k}_1-{\bf k}_4}\delta_{{\bf k}_1+{\bf k}_2-{\bf k}_3-{\bf k}_4}\sum_{i=9}^{20}\sum_{j\in\langle i,j\rangle}^{d\to d}\Big(e^{i\bf{k}_1(\bm{r}_j-\bf{r}_i)}d_{j-{\bf k}_1}^\dagger d_{i-{\bf k}_2}^\dagger d_{i-{\bf k}_3}d_{i-{\bf k}_4}\!+e^{i\bf{k}_4(\bm{r}_j-\bf{r}_i)}d_{j-{\bf k}_1}^\dagger d_{j\!-\!{\bf k}_2}^\dagger d_{j-{\bf k}_3}d_{i-{\bf k}_4}\\
&\hspace{25.6em}-2e^{i(\bf{k}_1-\bf{k}_4)\cdot(\bm{r}_j-\bf{r}_i)}d_{i-{\bf k}_1}^\dagger d_{j-{\bf k}_2}^\dagger d_{j-{\bf k}_3} d_{i-{\bf k}_4}\Big)\\
H_{ad}^{(4)}\!\!=&\frac{J_{ad}}{2N}\sum_{{\bf k}_1-{\bf k}_4}\delta_{{\bf k}_1+{\bf k}_2-{\bf k}_3-{\bf k}_4}\Bigg[\sum_{i=1}^8\sum_{j\in\langle i,j\rangle}^{a\to d}
\Big(e^{i\bf{k}_1(\bm{r}_j-\bf{r}_i)}a_{i{\bf k}_4}^\dagger a_{i{\bf k}_3}^\dagger a_{i{\bf k}_2} d_{j-{\bf k}_1}^\dagger+ e^{i\bf{k}_4(\bm{r}_j-\bf{r}_i)} d_{j-{\bf k}_1}^\dagger d_{j-{\bf k}_2}^\dagger d_{j-{\bf k}_3}a_{i{\bf k}_4}^\dagger\Big)\\
&\hspace{10em}+\sum_{i=9}^{20}\sum_{j\in\langle i,j\rangle}^{d\to a}\Big(e^{i\bf{k}_1(\bm{r}_j-\bf{r}_i)}a_{j{\bf k}_1}d_{i-{\bf k}_2}^\dagger d_{i-{\bf k}_3} d_{i-{\bf k}_4}+e^{i\bf{k}_4(\bm{r}_j-\bf{r}_i)}d_{i-{\bf k}_4}a_{j{\bf k}_3}^\dagger a_{j{\bf k}_2} a_{j{\bf k}_1}\\
&\hspace{26.6em}+4e^{i(\bf{k}_1-\bf{k}_4)\cdot(\bm{r}_j-\bf{r}_i)}a_{j{\bf k}_4}^\dagger a_{j{\bf k}_1} d_{i-{\bf k}_2}^\dagger d_{i-{\bf k}_3}\Big)\Bigg].
\end{aligned}\end{equation}
With the new notes of 
$\bm{x}_{\bf{k}}^\dag=\big(x_{1\bf{k}}^\dag \cdots x_{8\bf{k}}^\dag \; x_{9\bf{k}} \cdots x_{20\bf{k}}\big)=\big(a_{1\bf{k}}^\dag \cdots a_{8\bf{k}}^\dag\; d_{{9-\bf{k}}}\cdots d_{{20-\bf{k}}}\big)$,
$\bm{\chi}_{\bf{k}}^\dag=\big(\chi_{1\bf{k}}^\dag \cdots \chi_{8\bf{k}}^\dag \; \chi_{9\bf{k}} \cdots \chi_{20\bf{k}}\big)=\big({\alpha}_{1\bf{k}}^\dag \cdots {\alpha}_{8\bf{k}}^\dag\; {\beta}_{{1-\bf{k}}}\cdots {\beta}_{{12-\bf{k}}}\big)$, the Bogoliubov transformation is expressed as:
\begin{equation}
x_{i{\bf k}}=\sum_{m=1}^{20}\mathcal P_{\bf k}^{i,m}\chi_{m{\bf k}},\quad x_{i{\bf k}}^\dagger=\sum_{m=1}^{20}\left(\mathcal P_{\bf k}^{i,m}\right)^*\chi_{m{\bf k}}^\dagger,
\end{equation}
which has been suggested in Eq. (\ref{Diag}) in the main text. Substituting the Bogoliubov transformation into Eq. (\ref{HH4app}), $H^{(4)}$ can be expressed in terms of $(\chi_{m{\bf k}}^\dag\;\chi_{m{\bf k}})$. Then, the mean-field approximation is applied to derive the four-operator products into two-operator products. Within $H^{(4)}$, three distinct types of four-operator products are identified. The mean-field approximation pertaining to these three types of four-operator products is detailed as follows:
\begin{align}
\chi^\dag_{m_4\bf{k}_4}\chi^\dag_{m_3\bf{k}_3}\chi_{m_2\bf{k}_2}\chi_{m_1\bf{k}_1}\approx&(\delta_{\bf k_2\bf k_3}\delta_{m_2m_3}\delta_{m_1m_4}+\delta_{\bf k_1\bf k_3}\delta_{m_1m_3}\delta_{m_2m_4})\nonumber\\
&\times\Big[ (n_{m_2\bf{k}_2}+\xi_{m_2}^{\beta})\chi^\dag_{m_1\bf{k}_1}\chi_{m_1\bf{k}_1}+(n_{m_1\bf{k}_1}+\xi_{m_1}^{\beta})\chi^\dag_{m_2\bf{k}_2}\chi_{m_2\bf{k}_2}\Big],\label{chi-1}\\
\chi_{m_1\bf{k}_1}\chi_{m_2\bf{k}_2}\chi_{m_3\bf{k}_3}^\dag\chi_{m_4\bf{k}_4}^\dag\approx&(\delta_{\bf k_2\bf k_3}\delta_{m_2m_3}\delta_{m_1m_4}+\delta_{\bf k_1\bf k_3}\delta_{m_1m_3}\delta_{m_2m_4})\nonumber\\
&\times\Big[ (n_{m_2\bf{k}_2}+\xi_{m_2}^{\alpha})\chi_{m_1\bf{k}_1}\chi_{m_1\bf{k}_1}^\dag+ (n_{m_1\bf{k}_1}+\xi_{m_1}^{\alpha})\chi_{m_2\bf{k}_2}\chi_{m_2\bf{k}_2}^\dag\Big],\label{chi-2}\\
\chi^\dag_{m_4\bf{k}_4}\chi_{m_1\bf{k}_1}\chi_{m_2\bf{k}_2}\chi^\dag_{m_3\bf{k}_3}\approx&(\delta_{\bf k_2\bf k_3}\delta_{m_2m_3}\delta_{m_1m_4}+\delta_{\bf k_1\bf k_3}\delta_{m_1m_3}\delta_{m_2m_4})\nonumber\\
&\times\Big[ (n_{m_2\bf{k}_2}+\xi_{m_2}^{\alpha})\chi^\dag_{m_1\bf{k}_1}\chi_{m_1\bf{k}_1}+ (n_{m_1\bf{k}_1}+\xi_{m_1}^{\beta})\chi_{m_2\bf{k}_2}\chi_{m_2\bf{k}_2}^\dag\Big].\label{chi-3}
\end{align}
The $ n_{m\bf{k}}$ refers to the Bose distribution defined in Eq. (\ref{bose}). $\xi_{m}^{\beta}=1$ when $m\in\{9,10,\cdots,20\}$ and $\xi_{m}^{\alpha}$ =1 when $m\in\{1,2,\cdots,8\}$. Otherwise, $\xi_{m}^{\beta}$ and $\xi_{m}^{\alpha}$ are equal to zero. The four-operator product in Eq. (\ref{chi-1}) is derived from the three terms in $H_{aa}^{(4)}$ as well as the first and fourth terms in $H_{ad}^{(4)}$. During the mean-field approximation, thermal averages such as $\langle\chi_{m_4{\bf k}_4}^\dag\chi_{m_1{\bf k}_1}\rangle$ arise, and within these averages, the term $\langle\beta_{m_4-{\bf k}_4}\beta_{m_1-{\bf k}_1}^\dag\rangle$ exists. Consequently, the factor  $\xi_{m_1}^\beta$ emerges. Moreover, the four-operator product in Eq. (\ref{chi-2}) is derived from the three terms in $H_{dd}^{(4)}$ as well as the second and third terms in $H_{ad}^{(4)}$, and the four-operator product in Eq. (\ref{chi-3}) originates solely from the fifth term in $H_{ad}^{(4)}$. Due to a similar rationale that arises during the mean-field approximation process of Eq. (\ref{chi-1}), the factors $\xi_{m}^{\alpha}$, $\xi_{m}^{\beta}$ also manifest themselves in Eqs. (\ref{chi-2}) and (\ref{chi-3}).

Combining the results of operator parts with the vertex parts which contain the products of $\mathcal P_{\bf k}^{i,m}$'s and structure factors $e^{-i{{\bf k}_2}{\bf r}_{ij}}$, we can get the expressions of $H^{(4)}_{aa}$, $H^{(4)}_{dd}$ and $H^{(4)}_{ad}$ after the mean-field approximation. Here is a summary of them:
\begin{equation}\begin{aligned}
H^{(4)}_{aa}=&\frac{J_{aa}}{N}\sum_{{\bf k},m_1}\sum_{{\bf p},m_2}\sum_{i=1}^8\sum_{j\in\langle i,j\rangle}^{a\to a}(n_{m_2\bf{p}}+\xi_{m_2}^{\beta})\chi^\dag_{m_1\bf{k}}\chi_{m_1\bf{k}}\times M^{{\rm MultP}}_1,\\
H^{(4)}_{dd}=&\frac{J_{dd}}{N}\sum_{{\bf k},m_1}\sum_{{\bf p},m_2}\sum_{i=9}^{20}\sum_{j\in\langle i,j\rangle}^{d\to d}(n_{m_2\bf{p}}+\xi_{m_2}^{\alpha})\chi^\dag_{m_1\bf{k}}\chi_{m_1\bf{k}}\times M^{{\rm MultP}}_1,\\
H^{(4)}_{ad}=&\frac{J_{ad}}{N}\Big[\sum_{{\bf k},m_1}\sum_{{\bf p},m_2}\sum_{i=1}^8\sum_{j\in\langle i,j\rangle}^{a\to d}(n_{m_2\bf{p}}+\xi_{m_2}^{\beta})\chi^\dag_{m_1\bf{k}}\chi_{m_1\bf{k}}\times M^{{\rm MultP}}_2\\
&\quad+\sum_{{\bf k},m_1}\sum_{{\bf p},m_2}\sum_{i=9}^{20}\sum_{j\in\langle i,j\rangle}^{d\to a}(n_{m_2,\bf{p}}+\xi_{m_2}^{\alpha})\chi^
\dag_{m_1\bf{k}}\chi_{m_1\bf{k}}\times M^{{\rm MultP}}_2\Big],\\
\end{aligned}\end{equation}where\begin{equation}\begin{aligned}
M^{{\rm MultP}}_1=&2{\rm Re}\Bigg[\Big(\big|{\mathcal P}_{\bf p}^{i,m_2}\big|^2-e^{-i{\bf p}\cdot{\bf r}_{ij}}{\mathcal P}_{\bf p}^{i,m_2}{\mathcal P}_{\bf p}^{j,m_2*}\Big)\Big(e^{i{\bf k}\cdot{\bf r}_{ij}}{\mathcal P}_{\bf k}^{j,m_1}{\mathcal P}_{\bf k}^{i,m_1*}-\big|{\mathcal P}_{\bf k}^{j,m_1}\big|^2\Big)\Bigg]\\
M^{{\rm MultP}}_2=&2{\rm Re}\Bigg[\Big(\big|{\mathcal P}_{\bf p}^{i,m_2}\big|^2+e^{-i{\bf p}\cdot{\bf r}_{ij}}{\mathcal P}_{\bf p}^{i,m_2}{\mathcal P}_{\bf p}^{j,m_2*}\Big)\Big(e^{i{\bf k}\cdot{\bf r}_{ij}}{\mathcal P}_{\bf k}^{j,m_1}{\mathcal P}_{\bf k}^{i,m_1*}+\big|{\mathcal P}_{\bf k}^{j,m_1}\big|^2\Big)\Bigg]\\
\end{aligned}\end{equation}
By combining the three terms in equation (A9) and separating the parts that sum over momentum {\bf p} from the parts that sum over {\bf k}, we can obtain Eqs. (\ref{MF}), (\ref{delta}), and (\ref{eta}) in the main text.

\section{\MakeUppercase{The magnon current and energy current}} 
\label{app: B}
\renewcommand{\theequation}{B\arabic{equation}}  
\setcounter{equation}{0} 
Combining the continuity equation and Heisenberg's equation of motion, the expression of spin current can be derived as \cite{mahan2013many}
\begin{eqnarray}\label{JJSE}
\bm{J_S}=\sum_{n=1}^N\sum_{i=1}^{20}{\bf R}_{n,i}[H,S_{n,i}^z],
\end{eqnarray}
where $N$ denotes the total number of unit cells, $S_{n,i}^z$ and ${\bf R}_{n,i}$ denote the $z$-component of spin and the position vector of the $i$th site in the $n$th unit cell, respectively. The Hamiltonian $H$ can be approximately substituted by the harmonic Hamiltonian (\ref{HH2}).
Then the expressions of spin current in YIG are obtained in the real space as:
\begin{equation}\begin{aligned}
\bm{J_S}=i\sum_{n=1}^N\sum_{i=1}^{20}{\bf R}_{n,i}[H,S_{n,i}^z]\approx i\sum_{n=1}^N\sum_{i=1}^{8}{\bf R}_{n,i}[H^{(2)},S-a_{{\bf R}_{n,i}}^\dagger a_{{\bf R}_{n,i}}]+i\sum_{n=1}^N\sum_{i=9}^{20}{\bf R}_{n,i}[H^{(2)},-S+d_{{\bf R}_{n,i}}^\dagger d_{{\bf R}_{n,i}}].
\end{aligned}\end{equation}
After some derivation of commutation, the expression of spin current is obtained,
\begin{equation}\begin{aligned}
\bm{J_S}=i\sum_{n=1}^N\Bigg\{&\sum_{i=1}^{8}2S\Big[J_{aa}\sum_{j\in\langle i,j\rangle}^{a\to a}{\bf{R}}_{n,i}\Big(a_{{\bf{R}}_{n,i}+{\bf r}_{ij}}^\dag a_{{\bf{R}}_{n,i}}-a_{{\bf{R}}_{n,i}}^\dag a_{{\bf{R}}_{n,i}+{\bf r}_{ij}}\Big)\\&\hspace{2.3em}+J_{ad}\sum_{j\in\langle i,j\rangle}^{a\to d}{\bf{R}}_{n,i}\Big(d_{{\bf{R}}_{n,i}+{\bf r}_{ij}}a_{{\bf R}_{n,i}}\!\!-\!a_{{\bf R}_{n,i}}^\dagger d_{{\bf{R}}_{n,i}+{\bf r}_{ij}}^\dagger\Big)\Big]\\
-&\sum_{n=9}^{20}2S\Big[J_{dd}\sum_{j\in\langle i,j\rangle}^{d\to d}{\bf{R}}_{n,i}\Big(d_{{\bf{R}}_{n,i}+{\bf r}_{ij}}^\dagger d_{{\bf{R}}_{n,i}}-d_{{\bf{R}}_{n,i}}^\dagger d_{{\bf{R}}_{n,i}+{\bf r}_{ij}}\Big)\\
&\hspace{2.3em}-J_{ad}\sum_{j\in\langle i,j\rangle}^{d\to a}{\bf{R}}_{n,i}\Big(a_{{\bf{R}}_{n,i}+{\bf r}_{ij}}d_{{\bf{R}}_{n,i}}-d_{{\bf{R}}_{n,i}}^\dagger a_{{\bf{R}}_{n,i}+{\bf r}_{ij}}^\dagger\Big)\Big]\Bigg\}.
\end{aligned}\end{equation}
After the Fourier transformation, we obtain:
\begin{equation}\begin{aligned}
\bm{J_S}=&i2S\sum_{\bf k}\Bigg[\sum_{i=1}^{8}\bigg(J_{aa}\sum_{j\in\langle i,j\rangle}^{a\to a}\frac{\partial{e^{i\bf k\cdot{\bf r}_{ij}}}}{\partial {\bf k}}a_{i\bf k}^\dagger a_{j\bf k}+J_{ad}\sum_{j\in\langle i,j\rangle}^{a\to d}\frac{\partial{e^{i\bf k\cdot{\bf r}_{ij}}}}{\partial {\bf k}}a_{i\bf k}^\dagger d^\dag_{j\bf k}\bigg)\\
&\hspace{4em}+\sum_{i=9}^{20}\bigg(J_{dd}\sum_{j\in\langle i,j\rangle}^{d\to d}\frac{\partial{e^{i\bf k\cdot{\bf r}_{ij}}}}{\partial {\bf k}}d_{i\bf k} d_{j\bf k}^\dagger+J_{ad}\sum_{j\in\langle i,j\rangle}^{d\to a}\frac{\partial{e^{i\bf k\cdot{\bf r}_{ij}}}}{\partial {\bf k}}d_{i\bf k} a_{j\bf k}\bigg)\Bigg].
\end{aligned}\end{equation}
Finally, the expression of $\bm{J_S}$ can be summarized with the help of the notations in Eq. (\ref{HH2}) as:
\begin{equation}\begin{aligned}
\bm{J_S}=&\sum_{\bf{k}} \begin{matrix}\big(\bm{a}_{\bf{k}}^\dag & \bm{d}_{-{\bf{k}}}\big)\end{matrix}
\frac{-\partial}{\partial{\bf k}}\Bigg[\begin{matrix}\bm A_{\bf{k}} & \bm B_{\bf{k}}\\ \bm B_{\bf{k}} ^\dag & \bm D_{\bf{k}}\end{matrix}\Bigg]
\Bigg(\begin{matrix}\bm{a}_{\bf{k}}\\\bm{d}_{-{\bf{k}}}^\dag\end{matrix}\Bigg)
=\sum_{\bf k}{\bm\chi}_{{\bf k}}^\dagger{\bm{\mathcal P}}_{\bf k}^\dag\frac{-\partial {\bm{\mathcal H}}_{\bf k}}{\partial {\bf k}}{\bm{\mathcal P}}_{\bf k}{\bm \chi}_{{\bf k}}.
\end{aligned}\end{equation}

The expression of spin current can be derived as:
\begin{equation}
\bm{J_E}=\sum_{n=1}^N\sum_{i=1}^{20}{\bf R}_{n,i}[H,h_{n,i}]
\end{equation}
where $h_{n,i}$ is the Hamiltonian on the $i$th site in the $n$th cell: $H^{(2)}=\sum_{n=1}^{N}\sum_{i=1}^{20}h_{n,i}$, with $\sum_{i=1}^{20}h_{n,i}=\sum_{i=1}^{8}h_{n,i}^a+\sum_{i=9}^{20}h_{n,i}^d$. The $a$ site Hamiltonian $h_{n,i}^a$ and the $d$ site Hamiltonian $h_{n,i}^d$ are given by: 
\begin{equation}\label{hhad}\begin{aligned}
&h_{n,i}^{a}\!=(16J_{aa}S-12J_{ad}S-g\mu_B B)\;a_{\mathbf{R}_{n,i}}^\dagger a_{\mathbf{R}_{n,i}}\!-2J_{aa}S\!\!\!\sum_{\mathbf{|r_{ij}|}=\rm{r_{aa}}}\!\!\!(a_{\mathbf{R}_{n,i}}^\dagger a_{\mathbf{R}_{n,i}\!+\!\mathbf{r}_{ij}}+{\rm{h.c}})-2J_{ad}S\!\!\!\sum_{\mathbf{|r_{ij}|}={\rm r}_{ad}}\!\!\!(a_{\mathbf{R}_{n,i}}^\dagger d_{\mathbf{R}_{n,i}\!+\!\mathbf{r}_{ij}}^\dagger+{\rm{h.c}}),\\
&h_{n,i}^{d}\!=(8J_{aa}S-8J_{ad}S+g\mu_B B)\;d_{\mathbf{R}_{n,i}}^\dagger d_{\mathbf{R}_{n,i}}\!-2J_{dd}S\!\!\!\sum_{\mathbf{|r_{ij}|}=\rm{r_{dd}}}\!\!\!(d_{\mathbf{R}_{n,i}}^\dagger d_{\mathbf{R}_{n,i}\!+\!\mathbf{r}_{ij}}+{\rm{h.c}})-2J_{ad}S\!\!\!\sum_{\mathbf{|r_{ij}|}={\rm r}_{ad}}\!\!\!(d_{\mathbf{R}_{n,i}}^\dagger a_{\mathbf{R}_{n,i}\!+\!\mathbf{r}_{ij}}^\dagger+{\rm{h.c}}).
\end{aligned}\end{equation}
The energy current could be divided into two parts:
\begin{equation}\begin{aligned}\bm{J_E}=i\sum_{n=1}^N\sum_{i=1}^{20}{\bf R}_{n,i}[H,h_{n,i}]\approx i\sum_{n=1}^N\sum_{i=1}^{8}{\bf R}_{n,i}[H^{(2)},h_{n,i}^{a}]+i\sum_{n=1}^N\sum_{i=9}^{20}{\bf R}_{n,i}[H^{(2)},h_{n,i}^{d}].\end{aligned}\end{equation}
After careful derivation, the energy currnt in the ${\bf k}$ space is obtained as:
\begin{equation}\label{AAA3}\begin{aligned}
\bm{J_E}=&2J_{aa}St_{JA}\sum_{\bf k}\sum_{j\in\langle i,j\rangle}^{a\to a}\left(\frac{\partial}{\partial \bf k}e^{i{\bf k}{\bm r}_{ij}}\right)a_{i\bf k}^\dagger a_{j \bf k}
+J_{ad}S(t_{JA}-t_{JD})\sum_{\bf k}\sum_{i=1}^8\sum_{j\in\langle i,j\rangle}^{a\to d}\left(\frac\partial{\partial {\bf k}}e^{i{\bf k}{\bm r}_{ij}}\right)a_{i\bf k}^\dagger d_{j\bf -k}^\dagger\\
&-2J_{dd}St_{JD}\sum_{\bf k}\sum_{j\in\langle i,j\rangle}^{d\to d}\left(\frac\partial{\partial {\bf k}}e^{i{\bf k}{\bm r}_{ij}}\right)d_{i\bf -k}d_{j\bf -k}^\dagger+J_{ad}S(t_{JA}-t_{JD})\sum_{\bf k}\sum_{i=9}^{20}\sum_{j\in\langle i,j\rangle}^{d\to a}\left(\frac\partial{\partial {\bf k}}e^{i{\bf k}{\bm r}_{ij}}\right)d_{i\bf -k}a_{j\bf k}\\
&+2J_{aa}^2S^2\sum_{\bf k}\sum_{i=1}^8\sum_{j\in\langle\langle i,j\rangle\rangle}^{a \to a \to a}\left(\frac\partial{\partial {\bf k}}e^{i{\bf k}{\bm r}_{ij}}\right)a_{i\bf k}^\dagger a_{j\bf k}
+2J_{ad}^2S^2\sum_{\bf k}\sum_{i=1}^8\sum_{j\in\langle\langle i,j\rangle\rangle}^{a \to d \to a}\left(-\frac\partial{\partial {\bf k}}e^{i{\bf k}{\bm r}_{ij}}\right)a_{i\bf k}^\dagger a_{j\bf k}\\
&+2J_{ad}^2S^2\sum_{\bf k}\sum_{i=9}^{20}\sum_{j\in\langle\langle i,j\rangle\rangle}^{d \to a \to d}\left(\frac\partial{\partial {\bf k}}e^{-i{\bf k}{\bm r}_{ij}}\right)d_{i\bf -k}^\dagger d_{j\bf -k}
+2J_{dd}^2S^2\sum_{i=9}^{20}\sum_{j\in\langle\langle i,j\rangle\rangle}^{d \to d \to d}\left(-\frac\partial{\partial {\bf k}}e^{-i{\bf k}{\bm r}_{ij}}\right)d_{i\bf -k}^\dagger d_{j\bf -k}\\
&+2J_{ad}J_{dd}S^2\sum_{\bf k}\sum_{i=1}^8\sum_{j\in\langle\langle i,j\rangle\rangle}^{a \to d \to d}\Bigg[\left(-\frac\partial{\partial \bf k}e^{i{\bf k}{\bm r}_{ij}}\right)a_{i\bf k}^\dagger d_{j\bf -k}^\dagger+\left(-\frac\partial{\partial \bf k}e^{-i{\bf k}{\bm r}_{ij}}\right)a_{i\bf k}d_{j\bf -k}\Bigg]\\
&+2J_{aa}J_{ad}S^2\sum_{\bf k}\sum_{i=1}^8\sum_{j\in\langle\langle i,j\rangle\rangle}^{a \to a \to d}\Bigg[\left(\frac\partial{\partial \bf k}e^{-i{\bf k}{\bm r}_{ij}}\right)a_{i\bf k}d_{j\bf -k}+\left(\frac\partial{\partial \bf k}e^{i{\bf k}{\bm r}_{ij}}\right)a_{i\bf k}^\dagger d_{j\bf -k}^\dagger\Bigg].
\end{aligned}\end{equation}
Where $t_{JA}=-(16J_{aa}S-12J_{ad}S-g\mu_B B)$, $t_{JD}=-(8J_{aa}S-8J_{ad}S+g\mu_B B)$. Here, new summation notations such as $\sum_{j\in\langle\langle i, j\rangle\rangle}^{d \rightarrow d \rightarrow a}$ appear. The subscript $j\in\langle\langle i, j\rangle\rangle$ indicates that there is an intermediate iron $l$ between the $j$th site and the $i$th site. And the superscript  $d \rightarrow d \rightarrow a$ indicates that the $i$th site ion and the intermediate ion $l$ are $d$-type iron ions, while the $j$th site ion is a $d$-type iron ion. The sites $i$ and $l$ are connected by the $J_{dd}$ bond, and the sites $l$ and $j$ are connected by the $J_{ad}$ bond. Other summation notations appear in Eq. (\ref{AAA3}) follow the same habits as $\sum_{j\in\langle\langle i, j\rangle\rangle}^{d \rightarrow d \rightarrow a}$. Finally, the energy current could be given as 
\begin{equation}\begin{aligned}
\bm{J_E}=\frac12&\sum_{\bf{k}} \begin{matrix}\big(\bm{a}_{\bf{k}}^\dag & \bm{d}_{-{\bf{k}}}\big)\end{matrix}
\frac{\partial}{\partial{\bf k}}\left(\Bigg[\begin{matrix}\bm A_{\bf{k}} & \bm B_{\bf{k}}\\ \bm B_{\bf{k}} ^\dag & \bm D_{\bf{k}}\end{matrix}\Bigg]
\begin{bmatrix}\bm I_{8\times8} & \bm 0 \\ \bm 0 & -\bm I_{12\times12}\end{bmatrix}
\Bigg[\begin{matrix}\bm A_{\bf{k}} & \bm B_{\bf{k}}\\ \bm B_{\bf{k}} ^\dag & \bm D_{\bf{k}}\end{matrix}\Bigg]\right)
\Bigg(\begin{matrix}\bm{a}_{\bf{k}}\\\bm{d}_{-{\bf{k}}}^\dag\end{matrix}\Bigg)
=\frac12 \sum_{\bf k}{\bm\chi}_{{\bf k}}^\dagger{\bm{\mathcal P}}_{\bf k}^\dag\frac{\partial [{\bm{\mathcal H}_{\bf k}{\bf g}\bm{\mathcal H}}_{\bf k}]}{\partial {\bf k}}{\bm{\mathcal P}}_{\bf k}{\bm \chi}_{{\bf k}}
\end{aligned}\end{equation}

\section{\MakeUppercase{The retarded current-current correlation functions}} 
\label{app: C}
\renewcommand{\theequation}{C\arabic{equation}}  
\setcounter{equation}{0} 

With the help of Kubo formulas, ${L}_{\mu\nu}$'s can be expressed in terms of the current-current correlation functions as
\begin{equation}
{L}_{\mu\nu}=\lim_{\omega\rightarrow0}\frac{\Phi_{\mu\nu}^R(\omega)-\Phi_{\mu\nu}^R(0)}{i\omega},
\end{equation}
where $\Phi_{\mu\nu}^R(\omega)=\Phi_{\mu\nu}^R(i\Omega\rightarrow\omega+i\delta)$ is the retarded current-current correlation function.
These three longitudinal current-current correlation functions are defined as 
\begin{eqnarray}\label{PhiJJ}
\Phi_{11}(i\Omega_n)&=&\int_0^\beta d\tau e^{i\Omega_n\tau}\frac1N\langle T_\tau J_S^x(\tau)J_S^x\rangle,\\
\Phi_{12}(i\Omega_n)&=&\int_0^\beta d\tau e^{i\Omega_n\tau}\frac1N\langle T_\tau J_S^x(\tau)J_E^x\rangle,\\
\Phi_{22}(i\Omega_n)&=&\int_0^\beta d\tau e^{i\Omega_n\tau}\frac1N\langle T_\tau J_E^x(\tau)J_E^x\rangle.
\end{eqnarray}

\end{widetext}


\end{document}